\documentclass[12pt]{article}
\def\beq{\begin{equation}}
\def\eeq{\end{equation}}
\def\ap#1#2#3 {Ann. Phys. (NY) {\bf#1} (19#2) #3}
\def\err#1#2#3 {{\it Erratum} {\bf#1} (19#2) #3}
\def\ib#1#2#3 {{\it ibid.} {\bf#1} (19#2) #3}
\def\ijmp#1#2#3 {Int. J. Mod. Phys. {\bf#1} (19#2) #3}
\def\jetp#1#2#3 {JETP Lett. {\bf#1} (19#2) #3}
\def\mpl#1#2#3 {Mod. Phys. Lett. {\bf#1} (19#2) #3}
\def\np#1#2#3 {Nucl. Phys. {\bf#1} (19#2) #3}
\def\pl#1#2#3 {Phys. Lett. {\bf#1} (19#2) #3}
\def\prep#1#2#3 {Phys. Rep. {\bf#1} (19#2) #3}
\def\prev#1#2#3 {Phys. Rev. {\bf#1} (19#2) #3}
\def\prl#1#2#3 {Phys. Rev. Lett. {\bf#1} (19#2) #3}
\def\sjnp#1#2#3 {Sov. J. Nucl. Phys. {\bf#1} (19#2) #3}
\def\spj#1#2#3 {Sov. Phys. JETP {\bf#1} (19#2) #3}
\def\spu#1#2#3 {Sov. Phys. Usp. {\bf#1} (19#2) #3}
\def\zp#1#2#3 {Zeit. Phys. {\bf#1} (19#2) #3}

\begin{document}
\begin{titlepage}
\begin{center}
{\Large \bf Theoretical Physics Institute \\
University of Minnesota \\}  \end{center}
\vspace{0.2in}
\begin{flushright}
TPI-MINN-01/32-T \\
UMN-TH-2016-01 \\
July 2001 \\
\end{flushright}
\vspace{0.3in}
\begin{center}
{\Large \bf  Semiclassical suppression of black hole production in
particle collisions
\\}
\vspace{0.2in}
{\bf M.B. Voloshin  \\ }
Theoretical Physics Institute, University of Minnesota, Minneapolis,
MN
55455 \\ and \\
Institute of Theoretical and Experimental Physics, Moscow, 117259
\\[0.2in]
\end{center}

\begin{abstract}
It is argued that the cross section for production of large black holes,
for which a semiclassical description is applicable, cannot be given by
the geometric area of the black hole horizon, as claimed recently in
the literature. Rather the production cross section in a few-particle
collision is suppressed by at least a factor $\exp (-I_E)$ with $I_E$
being the Gibbons-Hawking (Euclidean) action for the black hole. Thus
only essentially non-classical small black holes with mass of the order
of the Planck mass can possibly be produced in few-particle collisions
at trans-Planckian energies.
\end{abstract}
\end{titlepage}

Understanding nonperturbative effects of virtual and real black holes in
particle collisions is a long-standing theoretical challenge
\cite{hawking,zeldovich}. One particular problem related to such effects
is that of the cross section for production of black holes in particle
collisions at trans-Planckian energies\cite{acv,hooft,avv}. The intrigue
in this problem stems both from the general issue of `quantum gravity'
and from the not yet completely understood problem of production of
semiclassical field configurations in high-energy collisions of quantum
particles. The latter problem was extensively tackled in a non-gravity
setting in connection with possible baryon and lepton number violating
processes in multi-TeV particle collisions in the standard electroweak
model, as well as generally in connection with nonperturbative
multi-boson production in weakly coupled theories\cite{mv1} and also
with the catalysis of a false vacuum decay by particle
collisions\cite{mv2}.

Clearly, trans-Planckian collisions in the standard gravity theory can
only be of a purely `academic' interest in view of inaccessibility of
the relevant energy scale. The situation however changes in the recently
popular schemes with extra spatial dimensions of the space-time, with
the new dimensions having an unusually `large' size \cite{add}. In these
schemes the equivalent of the Planck scale for the multi-dimensional
gravity can be as low as in a TeV range, which suddenly brings the issue
of trans-Planckian collisions into relevance for the LHC and possible
other realistic future colliders\cite{arefeva}. Most recently it has
been claimed \cite{gt,dl} that (multi-dimensional) large black holes
should be copiously produced once the energy of colliding partons
sufficiently exceeds the effective Planck scale, and that in fact such
production can be a dominant process at LHC. This claim is based on the
estimate of the production cross section for a black hole with horizon
of a radius $r_H$ as given by the geometric area of the horizon:
$\sigma_{H} \sim \pi \, r^2_H$.  In particular the claimed cross section
grows as energy in a positive power (depending on the number of extra
dimensions).

The purpose of this paper is to argue that the probability of production
of a {\it large} black hole, i.e. with the mass satisfying the
semiclassical condition $M_H \gg M_{\rm Pl}$, is not given by the
geometrical cross section area, but rather is suppressed by the factor
$\exp (-I_E)$ where $I_E$ is the Gibbons-Hawking action\cite{gh} for the
black hole, $I_E \propto (M_H/M_{\rm Pl})^{(D-2)/(D-3)}$, with $D$ being
the total dimension of the space-time. It should be pointed out that the
suppression of production of large black holes does not contradict the
possibility\cite{acv,hooft} that production of black holes can be
prominent at trans-Planckian energies. Rather it implies that the cross
section might be unsuppressed for processes with production of only {\it
small} black holes with mass $M_H$ of the order of $M_{\rm Pl}$, for
which $I_E = O(1)$, and which cannot be treated semiclassically, if a
treatment of such objects as resonances is possible at all.

In the most part of the discussion here a normal four-dimensional 
gravity theory is understood, and the straightforward generalization to 
higher dimensions is described in the end. Two lines of reasoning will 
be presented: one based on the path integral expression for the 
transition amplitudes, and the other one based on 
statistical/thermodynamical considerations.

The process under discussion is of the type $few \to H $, where the
initial state contains few particles (including the case of just two
particles colliding), and $H$ stands for a black hole with mass $M_H \gg
M_{\rm Pl}$.  The specification ``few" for the number of particles
implies here that the number of particles $n$ is not considered as a
large parameter\footnote{So that factors like $n!$ cannot compete with
the semiclassical exponential terms. Once this assumption is
invalidated, at asymptotically large $n$ one gets into the standard
behavior of a classical collapse of large number of particles into a
black hole.}. The transition amplitude for this process is given by the
path integral
\beq
{\cal A}(few \to H)= \int_{few (t=-\infty)}^{H (t=+\infty)} \exp(i I [g,
\phi])\, {\cal D} \phi {\cal D} g
\label{pi}
\eeq
over all the field trajectories starting with incoming few particles in
the distant past and ending as an outgoing black hole at $t=+\infty$,
and where $I[g,\phi]$ is the action functional depending on
the metric $g$ and all the rest fields, generically denoted as $\phi$.
The probability then is given, up to non-exponential flux factors, by
\beq
P(few \to H)=\sum_{H} \, {\cal A}^\dagger {\cal A}~,
\label{prob}
\eeq
where the sum runs over the states of the black hole. For a large black
hole a semiclassical calculation is justified. Noticing that the expression (\ref{prob}) involves a configuration with a black hole existing over an infinite (Minkowski) time, the exponential part
of the probability can be found from the full classical action
for the black hole, described by the metric ${\bar g}$,
\beq
P(few \to H) \sim  \exp (i \,I[{\bar g}]).
\label{saddle}
\eeq
It should be pointed out that the saddle-point expression (\ref{saddle})
describes the {\it entire sum} over the states of the black hole.

The classical action in eq.(\ref{saddle}) is given by the
Gibbons-Hawking\cite{gh} formula, which for a non-charged black hole
with mass $M_H$ and the angular momentum $J$ reads as $I[{\bar g}]=i \,
I_E(M_H,J)$ with
\beq
I_E(M_H,J)=2\pi \, G \, M_H^2 \, \left ( 1+  {1 \over \sqrt{1-j^2}}
\right )~,
\label{gha}
\eeq
where $G$ is the Newton's constant, and $j=J/(G M_H^2)$ is the angular
momentum in units of its maximal possible value. From this expression
and eq.(\ref{saddle}) one concludes that the probability of production
of a large semiclassical black hole is necessarily exponentially
suppressed.  Moreover, the total probability of production of black holes with mass $M_H$ and with different angular momenta is dominated by the contribution of slowly rotating black holes.
Indeed, the summation over the partial waves with different $J$ is
Gaussian and the exponential factor is determined by small $J$  :
\beq
P\left (few \to H(M_H) \right ) = \sum_J (2J+1) P\left (few \to H(M_H,J)
\right ) \sim \exp \left ( - 4 \pi \, G \, M_H^2 \right )~.
\label{fla}
\eeq
In other words, the production of rapidly rotating black holes with $j 
\sim O(1)$, which is argued in Ref.\cite{gt} to be a typical process, is 
in fact even more heavily suppressed by the semiclassical exponent, while the typical angular momenta contributing to the sum (\ref{fla}) are $\langle J \rangle \sim \sqrt{G} \, M_H$.

It has been also argued\cite{gt} that the reason for the claimed large
cross section ``is connected with the rapid growth of the density of
black hole states at large mass". However pursuing this argument
quantitatively, leads in fact to the same suppression as in
eq.(\ref{fla}). Indeed, the ``density" (number) of states is determined
by the entropy $S_H=4 \pi \, G \, M_H^2$ of the black hole as ${\cal
N}=\exp(S_H)$. The total probability of production of the black hole
states (eq.(\ref{prob})) can then be written as
\beq
P(few \to H)=\sum_{H} \, \left | {\cal A}(few \to H) \right |^2 \sim
{\cal N} \, \left | {\cal A}(few \to H) \right |^2~.
\label{pstat}
\eeq
On the other hand, by the CPT symmetry the amplitude ${\cal A}(few \to
H)$ is related\cite{hooft2} to the amplitude of decay of {\it each}
state of the black hole into the considered state of ``few"
(anti)particles: $|{\cal A}(few \to H)|^2=|{\cal A}(H \to {\overline
{few}})|^2$. The probability of such decay can be estimated from the
black hole evaporation law with the temperature $T_H=1/(4\pi \, r_h)$:
\beq
P(H \to few) \sim |{\cal A}(H \to {\overline {few}})|^2 \sim \exp \left
( - \sum_i {E_i \over T_H} \right ) = \exp \left ( -{M_H \over T_H}
\right )~,
\label{evap}
\eeq
where $E_i$ are the energies of individual particles. Thus, using the
CPT reciprocity, the probability in eq.(\ref{pstat}) can be evaluated as
\beq
P(few \to H) \sim \exp \left ( S_H - {M_H \over T_H} \right ) = \exp
\left (- 4 \pi \, G \, M_H^2 \right )~,
\label{statres}
\eeq
with exactly the same exponential suppression as in eq.(\ref{fla}). This
agreement should come as no surprise, since the expression in
(\ref{statres}) contains the free energy $F_H=M_H-T_H \, S_H$, in
agreement with the general thermodynamic expression for the probability
as being given by $\exp(-F/T)$, and since the Euclidean space
calculation of the action\cite{gh} is what gives precisely $F/T$ in a
thermodynamic interpretation.

It should be noted, that the estimate (\ref{evap}) of the decay
probability from the evaporation of the black hole is not entirely
without a caveat. Namely the standard consideration of
evaporation\cite{hawking75}, leading to the Gibbs factor $\exp(-E_i/T)$
per each particle, neglects the back reaction of the radiated particles
on the black hole. In the process of decay into few particles the black
hole disappears, and the effects of back reaction should be quite
important. One might expect however, that these effects do not
drastically change the exponent, estimated from the evaporation formula.
Indeed, if the number of (``few") particles $n$ is a large number $n \gg
1$, the emission of each of these particles does not significantly
affect the mass of the remaining black hole. Thus one might expect that
the back reaction gives corrections to pre-exponent decreasing for large
$n$. Extrapolating this behavior down to small $n$ and eventually down
to $n=2$ may significantly change the pre-exponent in eq.(\ref{evap}),
but the back reaction effects are unlikely to compete with the large
exponential factor. Certainly, the agreement of the result from this
estimate with that from the path integral consideration can be
argued as a reasoning for such behavior.

One can also notice that the considered process $few \to H$ is only a
special case of a more general class of processes, where additional
particles are produced in association with the black hole: $few \to H+
(few)^\prime$. The described path integral reasoning however is readily
generalized to this case, and for a fixed mass $M_H$ results in the same
exponential suppression as in eq.(\ref{fla}). If, on the other hand, the
total energy of initial particles is fixed and one is interested in the
cross section of production of a black hole with {\it any} mass, it is
clear that the exponential factor favors production of essentially
quantum black holes with $M_H = O(M_{\rm Pl})$, and the excess energy
should be radiated away in the form of ordinary particles in the final
state. If those small quantum black holes retain at least some
nonperturbative features of the large classical ones, such processes are
undoubtedly of an immense interest and may lead to qualitatively new
phenomena, e.g. to violation of global quantum numbers (baryon, lepton,
etc.)\cite{zeldovich}.

Finally, one can readily generalize the suppression factor of
eq.(\ref{fla}) to a multi-dimensional case, relevant for the models with
extra dimensions, using the Gibbons-Hawking (Euclidean) action for a
black hole in a $D$ dimensional space-time\cite{mp}:
\beq
I_E={4\pi \, M_H \, r_H \over (D-2) \, (D-3)}= {4\pi \, M_H  \over
(D-2) \, (D-3)}\, \left [ {8 \over (D-2) \, \pi^{(D-3)/2}} \, \Gamma
\left ( {D-1 \over 2} \right ) \, G_D \, M_H \right ]^{1/(D-3)}
\label{multi}
\eeq
with $G_D$ being the $D$ dimensional Newton's constant. Thus in either
number of dimensions the cross section for production of large black
holes with mass $M_H \gg M_{\rm Pl}$ exhibits an exponential
suppression: $\sigma \sim \exp[-c(D)\, (M_H/M_{\rm Pl})^{(D-2)/(D-3)}]$,
with $c(D)$ being a positive dimensionless constant, depending on the
number of dimensions $D$.

I thank G. Gabadadze, N. Itzhaki, K. Olive, and A. Vainshtein for useful
discussions. This work is supported in part by DOE under the grant
number DE-FG02-94ER40823.


\begin{thebibliography}{99}
\bibitem{hawking}
S.W. Hawking, \prev{D14}{76}{2460}.
\bibitem{zeldovich}
Ya. B. Zel'dovich, ZhETF {\bf 72} (1977) 18.
\bibitem{acv}
D. Amati, M. Ciafalone, and G. Veneziano, \pl{B197}{87}{81}; \
\np{B403}{93}{703}.
\bibitem{hooft}
G. 't Hooft, \pl{B198}{87}{61};\ \np{B304}{88}{867}.
\bibitem{avv}
I. Ya. Aref'eva, K.S. Viswanathan, and I.V. Volovich,
\np{B452}{95}{346}, Err: \np{B462}{96}{613}.
\bibitem{mv1}
An account of this development with a list of referencies can be found
in: M.B. Voloshin, {\it Non-Perturbative Methods}, Proc. XXVII Int.
Conf. on High Energy Phys., Glasgow, 1994, Ed. P.J. Bussey and I.G.
Knowles, IOP Pub., Bristol, 1995; p.121.
\bibitem{mv2}
M.B. Voloshin, \prev{D49}{94}{2014}.
\bibitem{add}
N. Arkani-Hamed, S. Dimopoulos and G. Dvali, \pl{B 429}{98}{263}.
\bibitem{arefeva}
I.Ya. Aref'eva, Steklov Math. Inst. Report SMI-25-99, Nov 1999; \
[hep-th/9910269].
\bibitem{gt}
S.B. Giddings and S. Thomas, Report NSF-ITP-01-62, SU-ITP-01-30, June
2001;\ [hep-ph/0106219].
\bibitem{dl}
S. Dimopoulos and G. Landsberg, Report, June 2001; \ [hep-ph/0106295].
\bibitem{gh}
G.W. Gibbons and S.W. Hawking, \prev{D15}{77}{2752}.
\bibitem{hooft2}
G. 't Hooft, Nucl.Phys. B (Proc. Suppl.) {\bf 43} (1995), 1.
\bibitem{hawking75}
S.W. Hawking, Commun.Math.Phys. {\bf 43} (1975) 199.
\bibitem{mp}
R.C. Myers and M.J. Perry, \ap{172}{86}{304}.


\end{thebibliography}
\end{document}